\documentclass[10pt,twocolumn,twoside]{IEEEtran}
\usepackage{amsmath,amssymb}
\usepackage{graphicx}
\usepackage{cite}
\usepackage{color}
\usepackage{array,makecell}
\newcolumntype{M}[1]{>{\centering\arraybackslash}m{#1}} 
\usepackage{setspace}  

\usepackage[table]{xcolor}
\definecolor{A1bg}{HTML}{E8F4FF} 
\definecolor{A2bg}{HTML}{E9F7EF} 
\definecolor{A3bg}{HTML}{FFF6E5} 
\definecolor{A4bg}{HTML}{FDE9EF} 
\definecolor{A5bg}{HTML}{F3E8FF} 
\definecolor{MIMObg}{HTML}{E8F1FF} 
\definecolor{RISbg}{HTML}{EAF9EE}  
\definecolor{SIMbg}{HTML}{FFF4E6}  

\begin{document}
\title{In-Wave Computation Aided Stacked Intelligent Metasurfaces in Next-Generation Networks:  Challenges and Opportunities}

\author{Mengbing Liu,  \emph{Graduate Student Member, IEEE}, 
Chau Yuen, \emph{Fellow, IEEE}, Dusit Niyato, \emph{Fellow, IEEE},  \\Bruno Clerckx, \emph{Fellow, IEEE}, Lajos Hanzo, \emph{Life Fellow, IEEE}
\thanks{
The work of Chau Yuen was supported by MOE (Ministry of Education, Singapore), under MOE Tier 2 Award number T2EP50124-0032.
The work of Lajos Hanzo was supported by the Engineering and Physical Sciences Research Council (EPSRC) through the following grants: Platform for Driving Ultimate Connectivity (TITAN) under Grant EP/Y037243/1 and EP/X04047X/1; Robust and Reliable Quantum Computing (RoaRQ, EP/W032635/1); India-UK Intelligent Spectrum Innovation ICON UKRI-1859; PerCom (EP/X012301/1); EP/X01228X/1; EP/Y037243/1. (\it{Corresponding authors: Chau Yuen; Lajos Hanzo.})}
 \thanks{Mengbing Liu and Chau Yuen are with the School of Electrical and Electronic Engineering, Nanyang Technological University, Singapore. (emails: mengbing001@e.ntu.edu.sg, chau.yuen@ntu.edu.sg). Dusit Niyato is with the College of Computing and Data Science, Nanyang Technological University, Singapore 639798 (e-mail: dniyato@ntu.edu.sg). Bruno Clerckx is with the Department of Electrical and Electronic Engineering, Imperial College London, SW7 2AZ London, U.K. (e-mail: m.nerini20@imperial.ac.uk; b.clerckx@imperial.ac.uk). Lajos Hanzo is with the School of Electronics and Computer Science, University of Southampton, Southampton, U.K. (e-mail: lh@ecs.soton.ac.uk).}
}

\maketitle

\begin{abstract}
 
Stacked intelligent metasurfaces (SIMs) facilitate computation by cascaded programmable layers so that part of the signal processing can be performed in the wave domain during signal propagation, rather than solely after reception. This approach expands the controllable degrees of freedom and supports the joint design of communication, sensing, and computation with the potential for reduced energy usage, shorter end-to-end latency, and improved task execution. Despite these advances, research on the SIM concept is still at an early stage, with challenges in scalability, controllability, nonlinearity, and robustness.  
This article reviews the state-of-the-art of SIM research, including applications, functions, and characteristics. We also demonstrate their potential through case studies on neural-like analog inference and communication enhancement. Finally, the paper outlines open challenges and future directions toward establishing SIMs as a new signal processing paradigm for in-wave computation in next-generation (NG) networks.
 
\end{abstract}
 
\section{Introduction}
Next-generation (NG) wireless networks are expected to deliver ultra-low latency, high reliability and stringent energy efficiency while supporting ever-growing traffic and intelligence at the network edge. Conventional systems still follow a “transmit then compute” paradigm, where raw waveforms are carried end-to-end and processed only after reception. This ``transport-first'' pipeline pushes sampling, buffering, and baseband workloads to the network edges. As bandwidth and antenna arrays scale, it aggravates backhaul congestion and significantly increases both latency and energy consumption.

Motivated by these limitations, a growing trend is to perform part of the signal processing during propagation. In this context, task-oriented operations are executed directly in the wave domain, reducing reliance on analog-to-digital and digital-to-analog conversions and heavy baseband processing. From this perspective, radio-frequency (RF) waveforms are shaped by a programmable channel so that the received signal approximates the output of a desired  operator. In line with this view, reconfigurable intelligent surfaces (RISs) constitute a low-cost and scalable platform that leverages phase programming to facilitate resource allocation and link adaptation \cite{Pan2022an}.

 While RISs demonstrate the potential of programmable wave control, conventional diagonal RIS architectures are typically implemented as single passive layers with element-wise phase control. Such a structure provides diagonal phase adjustments and limits controllability. Stacked intelligent metasurfaces (SIMs) extend this paradigm by cascading layers into a volumetric reconfigurable medium. The multilayer configuration introduces inter-layer interactions that expand the degrees of freedom (DoF) and enable richer electromagnetic transformations, paving the way for wave-domain signal-processing pipelines tailored to the stringent latency and efficiency requirements of NG wireless networks \cite{An2025emerging}.

A growing body of work explores SIMs as physical-layer computing substrates in the wave-domain. Recent studies show that SIMs can emulate key linear algebraic operators during wave propagation, such as singular value decomposition (SVD)-based transmit precoding (TPC) and zero-forcing (ZF) or minimum mean square error (MMSE) receive combining, to enable beamforming, interference suppression, and power allocation in multiuser links \cite{an2025stacked,papazafeiropoulos2025performance }. They can also perform frequency–angle transforms via fast Fourier transform (FFT) or discrete Fourier transform (DFT) for wideband structuring \cite{li2025fundamental}. Beyond these, SIMs support higher-level analog tasks for integrated sensing and communication (ISAC) and physical-layer security \cite{wang2024multi,niu2024efficient}. They also support diffractive computation for in-wave matrix operations, inference and localization \cite{liu2025onboard,javed2025sim}, which reflect a shift from passive reflection to programmable analog computing, where SIMs implement operators traditionally handled in the digital baseband. 
These emerging capabilities suggest that SIMs are playing increasingly diverse roles in wireless systems, which motivates the need for a systematic taxonomy to organize recent developments and to clarify their evolution toward NG wireless networks.

 \begin{table*}[t]
\begin{singlespace} 
\centering
\caption{Representative SIM  studies categorized by Scenarios (S), Characteristics (C), Functions (F), and grouped by Objectives (O), with Main Contributions.}
\label{tab:sim_acf_summary}
\renewcommand{\arraystretch}{1}
\setlength{\tabcolsep}{3.5 pt}
\footnotesize

\resizebox{\textwidth}{!}{%
\begin{tabular}{|M{3.2cm}|M{1 cm}|M{1.7cm}|M{1 cm}|p{7.9cm}|}
\Xhline{1.2pt}
 \textbf{Reference} & \textbf{Scenario} & \textbf{Characteristic} & \textbf{Function} & \makecell[c]{\textbf{Main Contribution}}\\
\Xhline{1.2pt}

\rowcolor{A1bg!55}\multicolumn{5}{|c|}{\textbf{O1: Communication enhancement}}\\ \hline
\rowcolor{A1bg!55} \makecell { J. An et al., \\ NTU, SG (2025) }& S1 & C1-C3 & F1 &
SIM realizes wave-domain multiuser downlink beamforming, offloading digital complexity and boosting data rate.\\ \hline 
\rowcolor{A1bg!55} \makecell{D. Darsena et al., \\ Univ. Naples Federico II, \\IT (2025)} & S1 & C1-C2 & F1 &
\vspace{-0.6em} Active SIM architecture delivers capacity gains over phase-only designs.\\ \hline 
\rowcolor{A1bg!55}\makecell{A. Papazafeiropoulos et al., \\ Univ. Hertfordshire, \\UK (2025)}
& S3 & C1-C3 & F1 &
\vspace{-0.6em} Double-SIM architecture improves uplink massive MIMO performance under statistical CSI.\\ \hline 
\rowcolor{A1bg!55}\makecell{Y. Zhang et al., \\Anhui Univ., CN (2025)}  & S1 & C1-C3 & F1 &
SIM enables uplink finite-blocklength communications with considerable sum-rate gains in Internet of Things systems.\\ \hline 
\rowcolor{A1bg!55}\makecell{H. Liu et al., \\ UESTC, CN (2025)}
& S1 & C1-C3 & F1 &
End-to-end wave-domain control of SIM enhances multiuser downlink rate.\\ \hline 
\rowcolor{A1bg!55}\makecell{Q. Huai et al., \\ ECUST, CN (2025)}
 & S1 & C1-C3 & F1 & SIM enhances rate-splitting multiple-access systems via wave-domain signal splitting and beamforming of common/private streams.\\ \hline 
\rowcolor{A1bg!55} \makecell{Z. Li et al., \\ NTU, SG (2025)} & S1 & C1-C3 & F2 &
Analog orthogonal frequency division multiple access with SIM reveals a fundamental resource–interference trade-off.\\ \hline 
\rowcolor{A1bg!55} \makecell{X. Yao et al., \\ UESTC, CN (2024)}
& S3 & C1-C3 & F3 &
SIM-tailored channel-estimation protocols resolve underdetermined holographic MIMO and reduce NMSE.\\ \hline 

\rowcolor{A2bg!55}\multicolumn{5}{|c|}{\textbf{O2: ISAC}}\\ \hline
\rowcolor{A2bg!55} \makecell{Z. Wang et al., \\ HUST, CN (2024)}
 & S4 & C1-C4 & F4 &SIM-aided ISAC validated by a three-layer 1-bit prototype with improved power and direction of arrival accuracy.\\ \hline 

\rowcolor{A3bg!55}\multicolumn{5}{|c|}{\textbf{O3: In-wave inference}}\\ \hline
\rowcolor{A3bg!55}  \makecell{M. Liu et al., \\ NTU, SG (2025)} & S1 & C1-C3, C5 & F5 &SIM-based linear inference architecture performs onboard terrain classification from raw Level-0 data, saving downlink bandwidth.\\ \hline 

\rowcolor{A4bg!55}\multicolumn{5}{|c|}{\textbf{O4: Security \& privacy}}\\ \hline
\rowcolor{A4bg!55} \makecell{H. Niu et al., \\ NTU, SG (2024) }& S1 & C1-C3 & F4 &Wave-domain joint modulation, beamforming, and artificial noise enable SIM-aided physical-layer security.\\ \hline 

\rowcolor{A5bg!55}\multicolumn{5}{|c|}{\textbf{O5: Positioning \& localization}}\\ \hline
\rowcolor{A5bg!55} \makecell{ A. Javed et al., \\ LUMS, PK (2025) }& S1 & C1-C3 & F3 &
A zone-based SIM fingerprinting framework enables indoor positioning with meter-level accuracy.\\ 
\Xhline{1.2pt}
\end{tabular}%
}
\end{singlespace}
\end{table*}
In this article, we establish a comprehensive taxonomy for SIMs that covers their scenarios, computational functions, characteristics, and system-level objectives. 
While existing SIM-related surveys mainly focus on channel enhancement and communication performance \cite{An2025emerging}, this article further explores SIMs from an in-wave computation perspective, with an emphasis on their computational functions in wireless systems. Building on this framework, we critically approach representative works to show how SIMs facilitate in-wave computation. We then analyze the benefits, design trade-offs, and challenges related to controllability, modeling, and algorithm–hardware co-design. Finally, we present illustrative case studies and outline inspirational directions for future research.
 
\section{Scenarios, Characteristics, Computational Functions, and  Objectives}

In this section, we review representative SIM-aided scenarios and analyze hardware and modeling characteristics \cite{liu2022programmable, Nerini2024Physically, An2025emerging, yahya2025t, Abrardo2025}. 
We then map these developments onto key computational functions and system objectives, primarily focusing on passive or quasi-passive SIM architectures unless otherwise specified, as summarized in Table~\ref{tab:sim_acf_summary}.

 \subsection{Scenarios (S)}
We categorize SIM-aided communication and sensing based on metasurface placement, which leads to four representative scenarios considered in existing studies.
\begin{itemize}
    \item \textbf{S1. SIM is located at the transmitter.}
In S1, the SIM is co-located with the transmitter and performs wave-domain  TPC, shaping the radiated field to suppress interference and reduce baseband overhead.
    \item \textbf{S2. SIM is located at the receiver.}
In S2, the SIM is co-located with the receiver and 
acts as a wave-domain receive combining right before the receiver to reduce reliance on digital baseband postprocessing.

    \item \textbf{S3. SIMs are located at both the transmitter and receiver.}
    In S3, SIMs are co-located at the transmitter and at the receiver, which are jointly configured as a cascaded programmable medium, increasing the  DoF, enabling complementary operations across the link, and reducing end-to-end latency and equalization burden.

    \item \textbf{S4. SIM is located in the middle.}
    In S4, the SIM is deployed midway along the propagation path and shapes the propagating wavefield to enhance coverage and link reliability, while enabling over-the-air wave-domain computation without changes to the transceivers.

\end{itemize}

\subsection{Characteristics (C)}
We analyze the properties that most directly impact controllability, latency–energy trade-offs, and learning workflows in SIM-aided systems as follows.
 \begin{figure*}[t]
    \centering
    \includegraphics[width=0.99\linewidth]{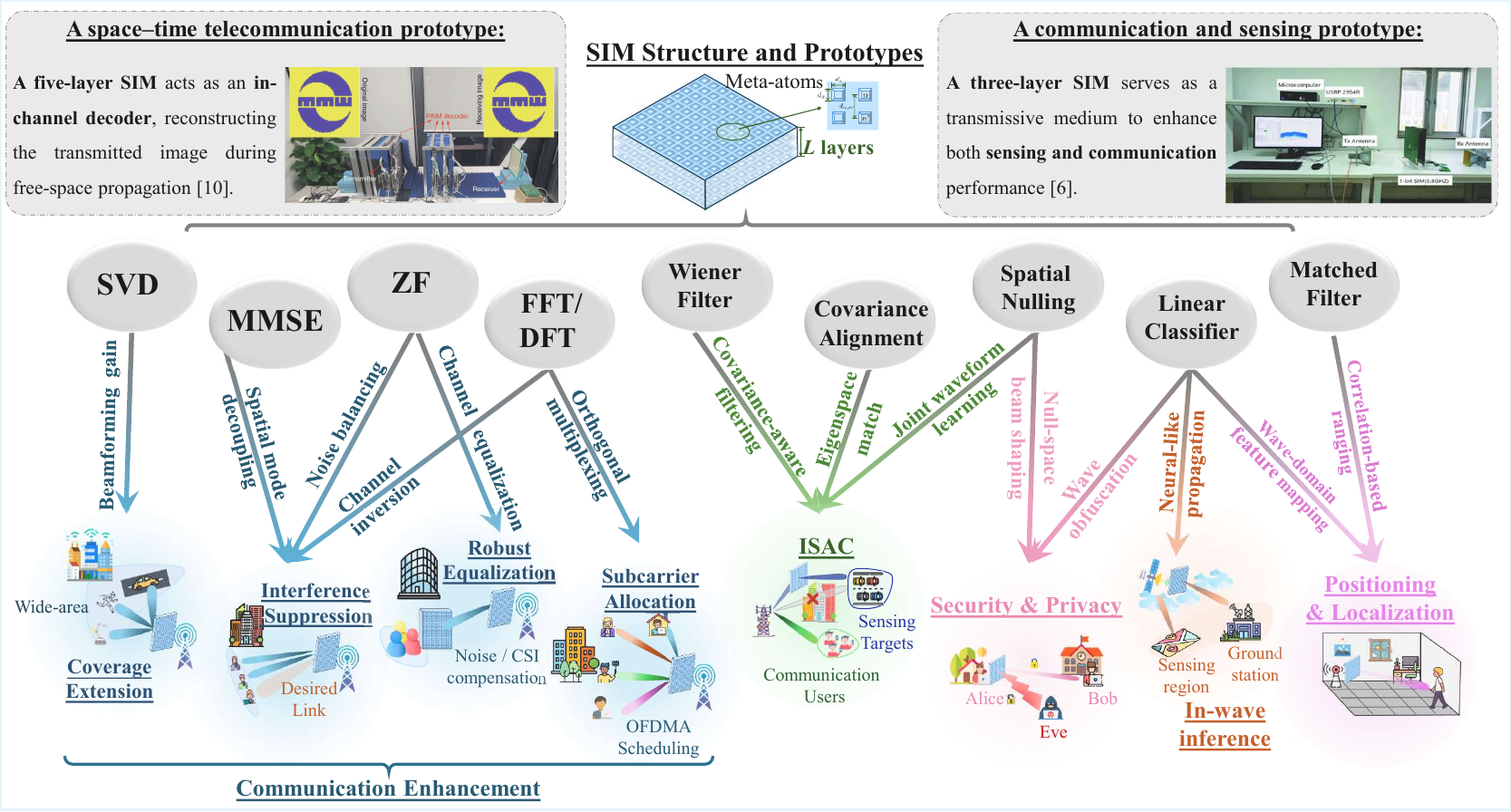}
    \caption{Overview of SIM computational functions and objectives. Functions are organized as wave-domain operators serving communication, sensing, and computation goals, with annotated links indicating supporting mechanisms. Two representative prototypes are shown: a five-layer programmable SIM acting as an in-channel decoder for space–time telecommunication \cite{liu2022programmable}, and a three-layer transmissive SIM enhancing communication and sensing performance \cite{wang2024multi}. }
    \label{fig:overview}
\end{figure*}
\begin{itemize}
    \item {\textbf{C1. Multilayer depth and geometry.}}
Stacking two or more programmable layers provides volumetric field control beyond what a single layer can achieve. By selecting the number of layers and customizing the design of each meta-atom, SIM supports richer transformations across successive layers and finer per-element control.

\item{\textbf{C2. Low-latency capability.}}
Executing part of the task within propagation reduces analog-to-digital or digital-to-analog conversion and baseband data movement, which lowers end-to-end delay and energy consumption \cite{An2025emerging}. Fast reconfiguration with lightweight online updates further supports time-sensitive operation and real-time services.

\item{\textbf{C3. Passive metasurface.}}
When implemented as a passive structure without RF chains or  amplification stages, SIM achieves field control through phase-only tuning at very low per-element power. This realization supports large apertures with high reliability and minimal cost.

\item{\textbf{C4. Hardware validation with quantization and switching.}}
SIM implementations offer discrete per-element control and fast reconfiguration.
Prototype measurements reveal practical imperfections that bound the control envelope and inform robust system design.

\item{\textbf{C5. Learning-native structure.}}
  Differentiable physics-based mappings represent per-layer phases as trainable parameters, allowing gradient-based end-to-end optimization of physical weights under task-specific loss functions, with passivity/quantization regularization and hardware-in-the-loop calibration. Compared to a digital learner that only outputs phase settings, this propagation-native approach reduces both dissipation and latency.
 
\end{itemize}

 \subsection{Computational Functions (F)}

We categorize the core wave-domain computations of  SIMs during propagation, where each function represents an analog operator for information transformation, combination, or inference. Fig.~\ref{fig:overview} shows the SIM structure and two representative prototypes: a five-layer in-channel decoding SIM and a three-layer transmissive SIM for joint sensing–communication. Representative operators are mapped to corresponding objectives, with annotated edges indicating their supporting mechanisms.

\begin{itemize}
    \item {\textbf{F1. TPC and receive combining.}}
    SIMs can implement linear projections that approximate a matrix pseudo-inverse required to achieve multiuser separation, interference suppression, and power allocation, as used in downlink multiuser and massive MIMO beamforming \cite{an2025stacked}.  
Operators include SVD, ZF, and MMSE, which are mimicked in SIMs by reproducing their 
linear projection behaviors at the waveform level through cascaded wave-domain interactions, rather than by explicit numerical computation.
They can realize linear algebraic functions required for beamforming, channel inversion, and interference coordination, making F1 primarily responsible for communication enhancement in SIM-aided systems.

    \item {\textbf{F2. Domain transformation.}}
SIMs can perform domain transforms for both separation and representation, such as FFT, DFT, and beamspace projection, which are widely used in orthogonal frequency division multiple access subcarrier mapping. 
Similarly, FFT/DFT transforms are mimicked in SIMs by structured multilayer wavefront shaping that approximates Fourier-like domain decompositions during propagation.
Such transformations support frequency- and angle-domain structuring for wideband resource allocation and domain mapping \cite{An2025emerging}. They also facilitate wavefront reparameterization under non-line-of-sight propagation conditions and dynamic spectrum shaping in adaptive or cognitive radio settings. For instance, F2 supports communication enhancement by improving orthogonality and exposure of usable subspaces, and it offers a potential path for integrated sensing by appropriately structuring covariance and feature spaces \cite{li2025fundamental}.

  \item {\textbf{F3. Correlation and detection.}}  
 SIM can realize inner-product correlation and matched filtering directly in the wave-domain to reveal target signatures and features, as in radar-style target detection and environment fingerprinting \cite{javed2025sim}. These operators facilitate analog detection, localization, and fingerprint identification without digital reconstruction, providing a low-latency and energy-efficient alternative to conventional sensing pipelines.

    \item {\textbf{F4. Conditioning and shaping.}}
 SIM can perform pre-equalization, spatial nulling, and covariance or eigenspace alignment to ameliorate the propagation channel. 
 For instance, Wiener filtering for MMSE equalization can be embedded as a wave-domain shaping objective to suppress interference and optimize second-order channel statistics.
 Beyond Wiener filtering,  equalization filters such as Kalman filters can be embedded as wave-domain shaping objectives to mitigate dispersion, suppress interference leakage, and optimize second-order statistics. F4  supports ISAC and security by shaping the environment, minimizing estimation uncertainty, and embedding physical-layer defenses through spatial angular filtering and artificial-noise injection \cite{niu2024efficient,wang2024multi}. 

\item {\textbf{F5. Neural-like analog computation.}}
SIM can execute cascaded linear projections, e.g., matrix multiplications, across programmable layers and produce compact analog readouts, enabling feature extraction and decision making during propagation, as in onboard SAR terrain classification \cite{liu2025onboard}. These computations realize linear inference and task-oriented compression, reduce reliance on digital postprocessing, and lower end-to-end latency and data transfer. F5   supports task-level decisions and computation offloading. When higher expressivity is required, it can be combined with controllable nonlinearity or digital refinement in a hybrid analog–digital design \cite{Gao2023SPNN, Sanjari2025NonlinearMetasurface}.

\end{itemize}

\subsection{Objectives (O)}

We organize the roles of SIMs into five objectives that reflect their system-level utility. Specifically, most existing research related to SIMs emphasizes communication enhancement, but its extension to  ISAC,  in-wave inference,  security \& privacy, and positioning \& localization is gaining popularity.

\begin{itemize}
 
  \item \textbf{O1. Communication enhancement.}
Through computational functions, SIMs enhance communication by executing key linear-algebraic and transform operations directly in the wave domain. Their primary role lies in  F1 precoding/combining, where SIMs implement analog projections to separate users, suppress interference, and allocate power,  avoiding digital beamforming and improving latency and energy efficiency \cite{an2025stacked, papazafeiropoulos2025performance}. For instance, SIM-based ZF TPC in the multiuser multiple-input single-output downlink can replace digital  TPCs and achieve up to 200\% sum-rate gain \cite{an2025stacked}.
Beyond F1, recent treatises explore  F2 transformation and F3 correlation functions. 
For instance, SIMs can structure the frequency–angle domain to orthogonalize subcarriers before scheduling, exposing a trade-off between spatial reuse and inter-user interference in orthogonal frequency division multiple access systems.
In such wideband settings, the frequency-dependent phase responses of metasurface elements may introduce beam squint effects that should be considered in practical SIM implementations \cite{li2025fundamental}.

\item \textbf{O2. ISAC.}
SIM-enabled ISAC is driven by conditioning and shaping functions in F4. In \cite{wang2024multi}, the SIM is optimized for aligning eigenspaces and controlling covariance in propagation, minimizing the estimation variance for direction of arrival, while satisfying per-user signal-to-interference-plus-noise ratio targets. Compared to the single-layer structure, a three-layer 1-bit prototype reports a 60\% reduction in direction of arrival estimation error and a $6$–$9$\,dB receive-power gain.

\item \textbf{O3. In-wave inference.}
 SIMs function as diffractive computers that execute in-wave inference in F5, where layered analog matrix multiplications and compact readouts perform feature extraction and decision mapping during propagation. For example, a satellite-borne SIM-based linear inference architecture carries out end-to-end supervised learning for terrain classification   from raw Level-0 in-phase/quadrature (I/Q) data,    reducing reliance on downlink bandwidth and ground computing, while attaining   90\% accuracy in land/ocean segmentation \cite{liu2025onboard}.
 
\item \textbf{O4. Security \& privacy.} 
SIM-enabled physical-layer security and privacy are driven by conditioning and shaping in F4. The SIM steers energy toward the legitimate receiver, while forming angular nulls toward eavesdroppers, realizing analog null-space projection and gain shaping during propagation. In \cite{niu2024efficient}, a closed-form iterative scheme updates per-layer phases and transmit power to maximize secrecy rate in single-input single-output links, achieving wave-domain confidentiality without cryptographic overhead. Simulation results demonstrate that the SIM-aided system achieves secure communication, while saving about 2\,dB of transmit power for a five-layer SIM.

\begin{table*}[t]

\centering
\caption{Computation-centric Comparison of MIMO, RIS, and SIM}
\renewcommand{\arraystretch}{1.12}
\setlength{\tabcolsep}{4pt}
\footnotesize
\begin{tabular}{|M{1.2cm}|M{2cm}|M{3cm}|M{2.8cm}|M{1.2cm}|M{1.2cm}|p{4.5cm}|}
\Xhline{1.2pt}  
\textbf{Scheme} & \textbf{Location} & \textbf{Operator richness} & \textbf{DoF} & \textbf{Speed} & \textbf{Energy} & \makecell[c]{\textbf{Computation roles}}\\[2pt]
\Xhline{1.2pt}

\rowcolor{MIMObg}
\textbf{MIMO} &
\makecell{Digital baseband}
&
Full digital algebra&
\makecell{High\\(Active antennas)} &
Slow &
High &
Digital matrix inversion, filtering, and beamforming. \\
\hline

\rowcolor{RISbg}
\textbf{RIS} &
wave-domain  &
Low-rank projection&
\makecell{Limited\\(single passive layer)}&
Fast &
Low &
Beam steering, coarse wave projection. \\
\hline

\rowcolor{SIMbg}
\textbf{SIM} &
wave-domain &
Layered analog operators&
\makecell{Very high\\(multiple passive layers)}
&
Fast&
Low &
Wave-domain computation (F1--F5). \\
\Xhline{1.2pt}
\end{tabular}
\label{tab:comparison_compute}
\end{table*}
 
 \item \textbf{O5. Indoor positioning \& localization.}
SIM-aided localization relies primarily on correlation and matched filtering in F3. By beneficially shaping the spatial RF power distribution and beamspace patterns, the SIM enhances fingerprint separability and carries out analog correlation during propagation. In \cite{javed2025sim}, a zone-based framework that combines alternating optimization of SIM phases with a lightweight $k$-nearest neighbors readout attains meter-level accuracy. Quantitatively, for an eight-zone design with a five-layer SIM, the mean localization errors are 1.72\,m with continuous-valued phase and 1.66\,m with discrete-valued phase, respectively.
 \end{itemize} 
 
 Research on SIM-aided systems has rapidly expanded, moving beyond communication objectives toward compute-in-propagation capabilities. 
 Table~\ref{tab:sim_acf_summary} organizes representative studies by scenario, objective, and computational function, highlighting SIM-aided analog operators built around math primitives and emerging roles in statistical conditioning and in-wave inference.
 These developments position SIMs as physical computing operators that execute core transformations during propagation, rather than devices that modify the environment. Despite this progress, SIM research presents compelling early opportunities to advance computation-driven signal processing in the wave domain, as discussed in the next section.

\section{Benefits, Challenges and Opportunities}

In this section, we compare SIMs with multiple-input multiple-output (MIMO) and RIS architectures from a computational perspective and identify the key design challenges and research opportunities for practical deployment.

\subsection{Benefits over MIMO and RIS}

Table~\ref{tab:comparison_compute} compares MIMO, RIS, and SIM techniques in terms of computation location, operator richness, spatial DoF, processing speed, energy cost\footnote{The energy cost characterizes the reliance on RF chains and energy-intensive baseband processing along the signal path (e.g., tens to hundreds of milliwatts per RF chain, depending on implementation), while processing speed reflects whether signal processing is performed during wave propagation or after full digital reception, which impacts end-to-end latency scaling, and their physical-layer computation roles from an architectural perspective.}.

\begin{itemize}
\item \textbf{MIMO.} For MIMO, computation resides in the digital baseband, typically executed on digital signal processors or field-programmable gate arrays. This enables a full set of linear operations with high numerical precision, supporting rich algebraic functions such as matrix inversion,  SVD, ZF, MMSE filtering,  DFT, and correlation.  These operators support flexible beamforming, spatial multiplexing, and interference cancellation in multiuser MIMO systems. However, their spatial DoF is bounded by the number of active RF chains, and the architectural flexibility comes at the cost of microsecond-to-millisecond digital latency and high energy dissipation.
As the system scale increases, centralized digital baseband processing in MIMO systems may face scalability challenges in terms of processing latency and energy consumption.

\item  \textbf{RIS.} RIS performs analog processing in the wave domain by adjusting single-layer phase configurations. Each element can be reconfigured with minimal energy and fast switching, enabling beam steering and coarse spatial shaping. However, its computational capability is limited to low-rank two-dimensional transformations, and the lack of a multilayer structure restricts its DoF and operator diversity   \cite{Pan2022an}. Thus, RIS is well suited to coverage enhancement and static channel manipulation, but not to complex or dynamic physical-layer tasks such as real-time equalization or joint sensing and communication \cite{An2025emerging}.

\item \textbf{SIM.}
  A SIM  transforms the propagation medium into a programmable analog computing substrate. Its multilayer architecture may realize a range of in-wave operators, such as pseudo-inverse projections for TPC and receive combining, FFT/DFT-based transforms for frequency–angle structuring, analog correlation and filtering for channel training, and statistical conditioning for robust inference \cite{an2025stacked,li2025fundamental, papazafeiropoulos2025performance, wang2024multi,liu2025onboard,niu2024efficient,javed2025sim}. These operations are performed natively during wave propagation at the speed of light and with low energy dissipation. 
SIMs also offer very high controllable DoF through volumetric manipulation, enabling near real-time waveform processing and supporting advanced objectives in communication, sensing, in-wave inference, security, and localization as O1-O5.  In this case, SIMs can function as a physical-layer pre-processing stage that reshapes the signal space prior to baseband processing,  alleviating the reliance on energy-intensive digital signal processing in MIMO systems.

\end{itemize}

Viewed from this perspective, SIMs shift core physical-layer operators from the digital baseband into the propagation domain, reducing computational load and energy consumption, while maintaining end-to-end system performance.

\subsection{Challenges and Opportunities}

Table~\ref{tab:comparison_compute} shows that SIMs enable richer wave-domain operations than conventional MIMO and RIS, while introducing unique challenges for scalable and reliable deployment.

\begin{figure*}[t]
    \centering
    \includegraphics[width= 0.95\linewidth]{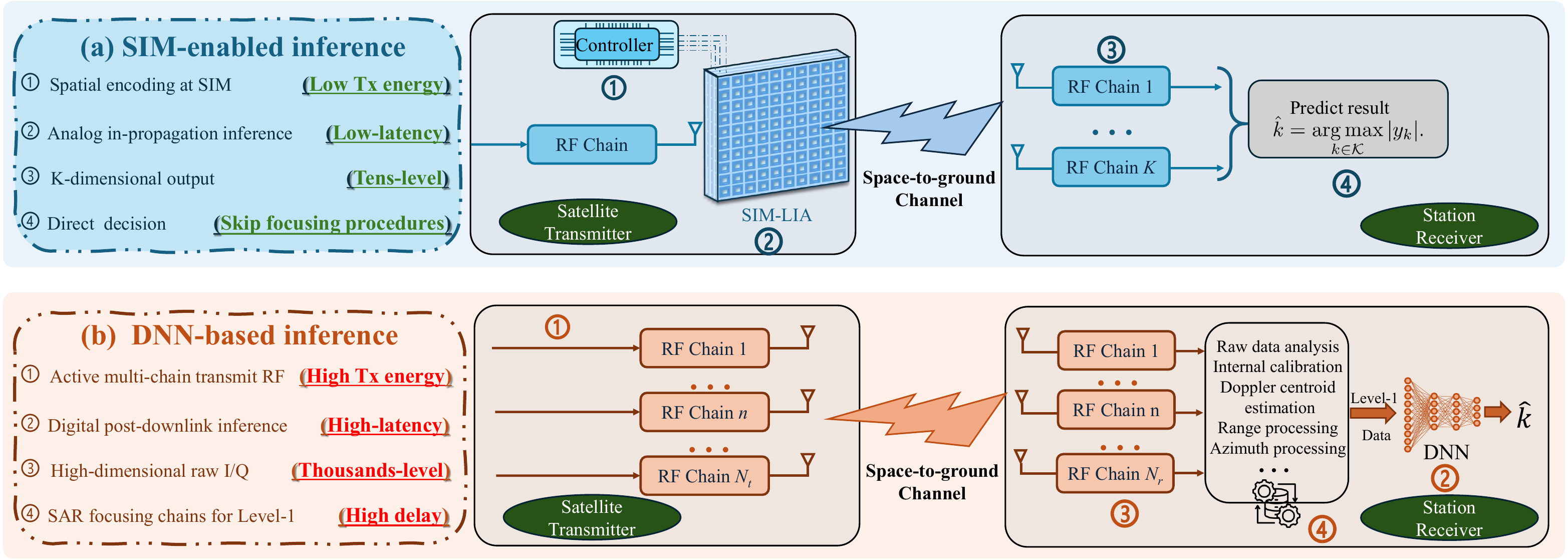}
    \caption{SIM-aided vs. digital inference over a satellite-to-ground link. 
(a) SIM-LIA performs spatial encoding and in-propagation inference, yielding a compact $K$-dimensional decision without Level-1 focusing. 
(b) The digital baseline transports high-dimensional I/Q for post-downlink processing and neural inference, increasing energy, latency, and bandwidth. }
    \label{fig:compare}
\end{figure*} 
 
\begin{itemize}
    \item {\textbf{Physical modeling and controllability.}}  SIMs enable volumetric wave manipulation with high electromagnetic DoF via stacked layers, but their performance is sensitive to electromagnetic and circuit-level modeling assumptions, including loss, dispersion, coupling, and phase quantization. 
 When realistic losses and inter-layer coupling are considered, increasing the number of SIM layers may exhibit diminishing returns. Under a fixed total number of elements, physically consistent $T$-parameters-based models show that a higher layer count reduces the per-layer aperture and accumulates transmission losses, thereby constraining the achievable sum-rate \cite{yahya2025t}. 
These effects limit the conditioning and controllability of the resultant operators. 
For example, finite-resolution phase control introduces quantization errors that degrade the conditioning of F1 transmit power control and receive combining, amplifying inversion noise and limiting interference suppression compared to ideal continuous-phase models \cite{An2025emerging}. 
Hence, constraint-aware and hardware-aligned modeling and optimization are essential for reliable and physically realizable SIM performance, and remain key to bridging the gap between idealized models and practical hardware implementations.
Beyond multilayer stacking, controllability can also be enhanced through per-layer topology design, as a single-layer beyond-diagonal RIS introduces intra-layer meta-atom interconnections that enrich the operator space \cite{Nerini2024Physically}.

\item {\textbf{Intrinsic linearity.}}  
Each programmable layer acts as a linear phase operator, and stacking preserves linearity. 
As a result, purely linear cascades have limited expressive power for inference-oriented tasks that require nonlinear decision boundaries or semantic compression, leading to a function-approximation bottleneck in in-wave inference. 
Since linear propagation limits the ability to realize nonlinear mappings in the wave domain \cite{liu2025onboard},  a practical direction is to introduce controllable nonlinearity and model it as activation constraints within the electromagnetic framework. 
Recent studies have explored this idea by inducing state-dependent responses through surface-programmable layouts \cite{Gao2023SPNN} or bias-programmable metasurface cells with nonlinear transfer characteristics \cite{Sanjari2025NonlinearMetasurface}. 
These approaches show that enhanced nonlinear expressivity can be achieved at the cost of increased circuit complexity, power consumption, and calibration overhead, highlighting a trade-off between representational capability and hardware complexity. 
From this perspective, hybrid analog–digital designs offer a pragmatic balance, while active SIM architectures with gain elements or active devices further extend the design space by enabling stronger and more controllable nonlinear responses through tunable biasing or active elements.

\item {\textbf{High-dimensional channel estimation.}}  
Accurate channel estimation is essential for reliable in-wave computation, as the metasurface must learn its effective channel before inference or beam control \cite{An2025emerging}. 
Direct estimation requires long pilot sequences and large regressions, with achievable NMSE limited by sample scarcity and ill-conditioned matrices. 
Structured training that aligns pilots with the Kronecker or beamspace channel structure improves efficiency and stability, while joint pilot–SIM optimization enhances identifiability with low complexity.

\item {\textbf{Robust in-wave computation under imperfect CSI.}} Frequent reconfiguration and channel variation make channel state information (CSI) mismatch unavoidable, so algorithms must remain stable under quantization, outdated statistics, and estimation errors. From a computational standpoint, imperfect CSI propagates through all processing functions and degrades both communication and inference performance. For instance, uncertainty degrades interference suppression in F1 TPC, while mismatched channel states reduce decision reliability in F5 inference.  
Robust computation, therefore, requires uncertainty-aware formulations that represent CSI variation through bounded or distributional sets and apply regularized inverses or condition-controlled updates.

 \begin{figure}[t]
  \centering
  \includegraphics[width= 0.95\linewidth]{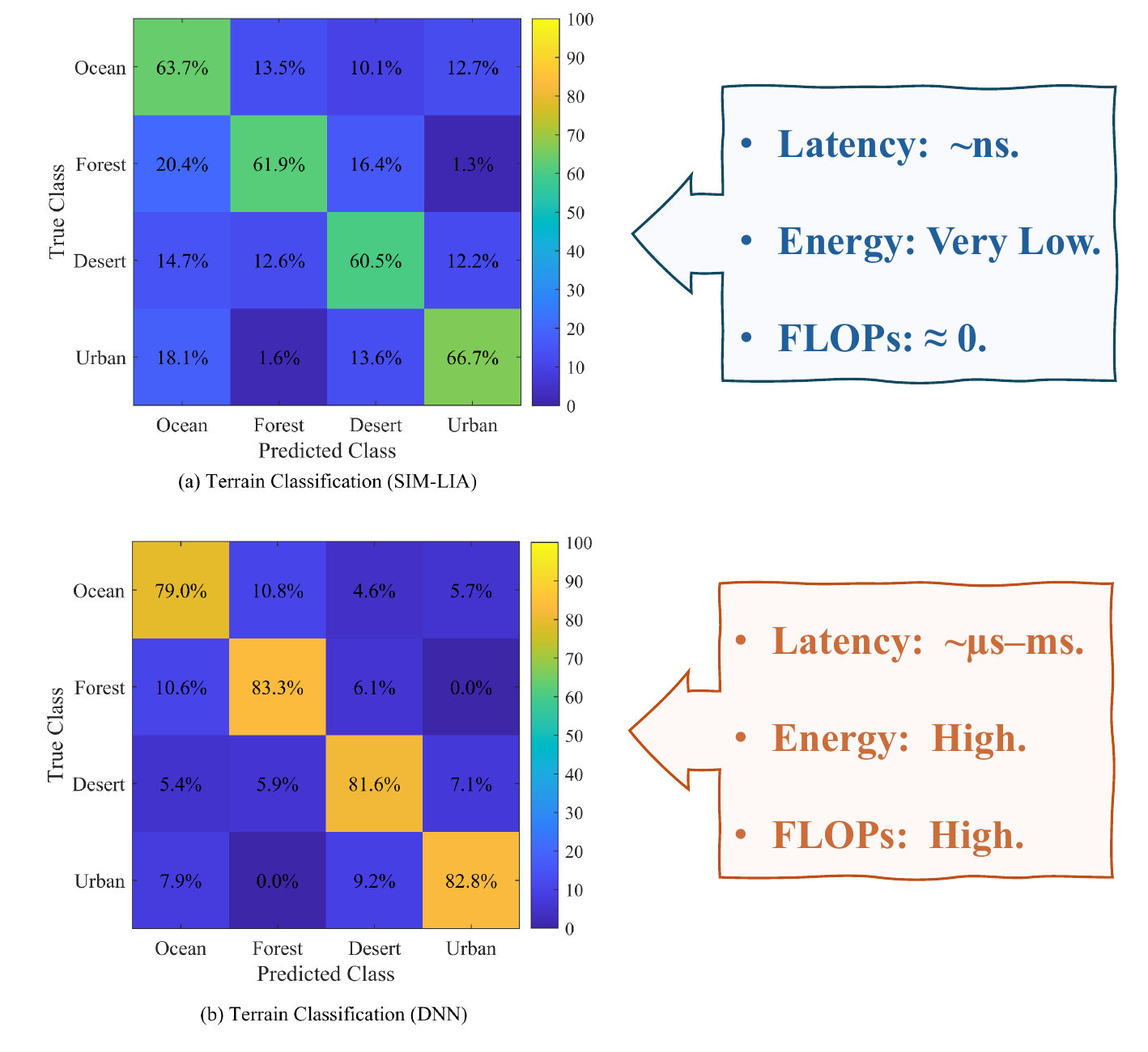}
  \caption{Confusion matrices of SIM-LIA and DNN. SIM-LIA offers comparable accuracy with lower latency and energy consumption.}
  \label{fig:case2_confmat}
\end{figure}

\item {\textbf{Data efficiency and security.}}  
Learning wave-domain operators requires supervision that links system states to task performance, but closed-loop measurements are scarce and costly, which limits the training of in-wave computations such as F2 transforms and F5 inference functions that rely on data-driven calibration. 
With limited training data, direct empirical risk minimization tends to overfit hardware-specific behaviors rather than learning mappings that generalize across SIM configurations. 
Self-supervised and unsupervised learning paradigms offer promising directions to alleviate data scarcity while preserving task-relevant structure. 
In parallel, physical-layer security approaches that exploit the high spatial DoF of SIMs, such as joint beamforming and artificial noise generation, offer a method to confine information to intended spatial modes \cite{niu2024efficient}. 
SIM-based in-wave computation is susceptible to configuration spoofing and waveform interception. Securing these systems demands robust control mechanisms and cross-layer security frameworks.

\end{itemize}

\section{In-wave neural-like computation for analog inference enabled by SIM}
 
This section examines SIM-aided analog inference and contrasts the SIM-LIA architecture with conventional digital deep neural network (DNN) pipelines, highlighting linear wave-domain computation versus nonlinear digital processing.

\subsection{Proposed Scheme: SIM-LIA versus Conventional DNN}

 \begin{figure}[t]
  \centering
  \includegraphics[width=0.95\linewidth]{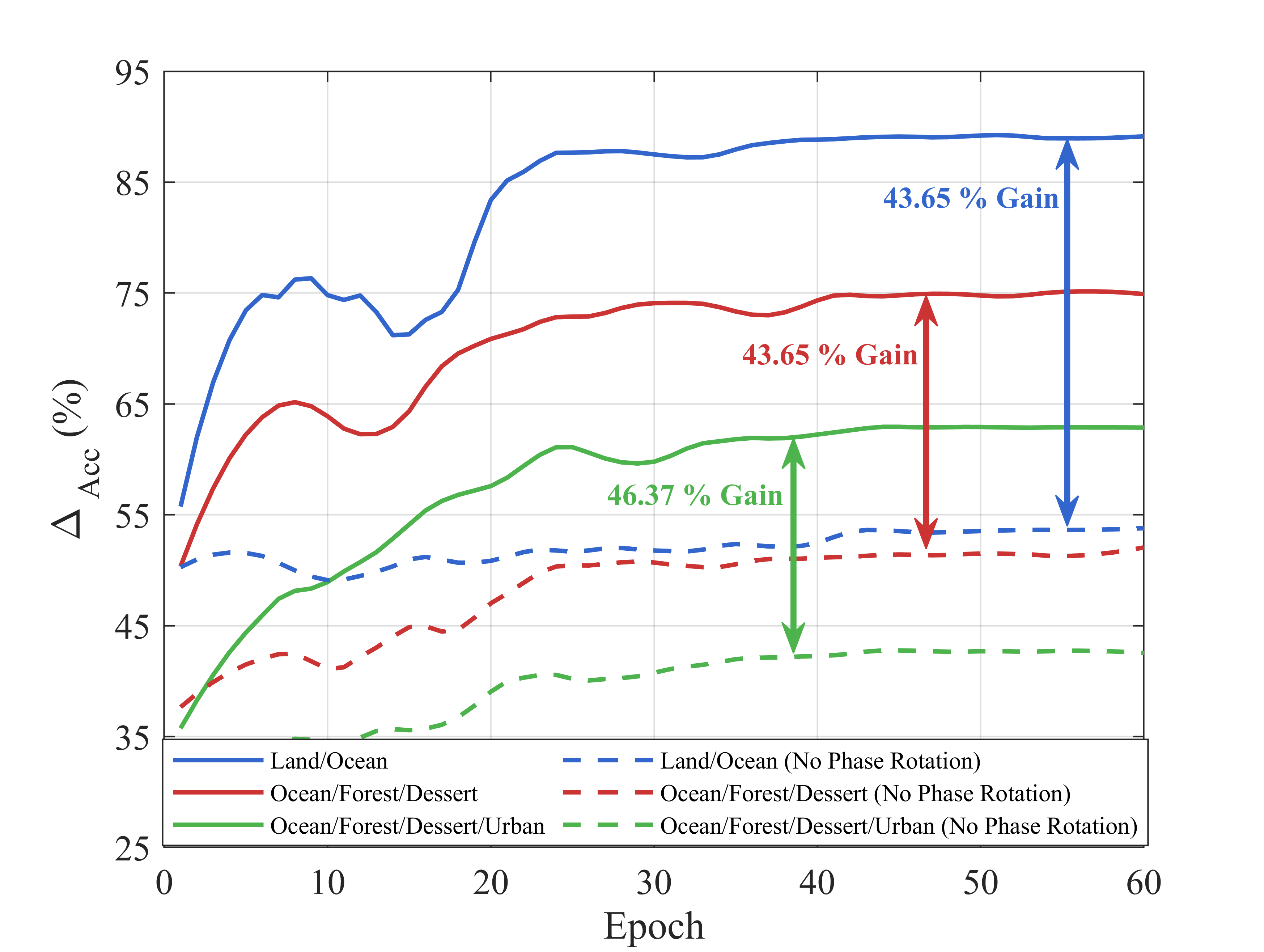}
  \caption{Epochs versus accuracy under different tasks. Wave-domain phase rotation consistently improves accuracy by an average of 45\%.}
  \label{fig:convergence}
\end{figure}
Fig.~\ref{fig:compare} contrasts two inference pipelines over a satellite-to-ground link, modeled as a dominant line-of-sight  Rician channel capturing free-space loss, atmospheric attenuation, and residual scattering.
In Fig.~\ref{fig:compare}(a), a small number of RF chains illuminate a multilayer SIM that performs spatial encoding at the transmitter and applies a learned diffractive transform, with phase configurations obtained offline and applied during inference\footnote{For clarity, a single transmit antenna is adopted to focus on the SIM-enabled in-wave computation under onboard constraints, such as power consumption and RF-chain complexity. Extensions to multi-antenna systems are considered a potential direction for future work.}.  
The ground station receives a $K$-element decision vector and makes the final decision without Level-1 focusing, where Level-1 data is a kind of high-level data after appropriate processing for easier inference. 
This placement reduces transmit energy, shortens end-to-end latency, and compresses the downlink payload. A shallow decision module is sufficient at the receiver, which preserves compatibility with standard processing chains.
In Fig.~\ref{fig:compare}(b), the digital baseline retains all operators after downlink. Multiple active RF chains transmit high-dimensional raw I/Q data to the ground, where calibration, motion compensation, range and azimuth processing, and Level-1 focusing precede neural inference. This design offers software flexibility and rich nonlinearity, but incurs higher energy consumption, processing delay, and bandwidth usage, since feature extraction begins only after full data reception and the dominant cost shifts to baseband processing.
Overall, Fig.~\ref{fig:compare} summarizes the trade-offs between in-medium feature extraction and software-based inference in terms of hardware reliance, expressivity, and processing cost.

\subsection{Task Description and Setup}
  
Building on the workflow contrast in Fig.~\ref{fig:compare}, we develop both pipelines for onboard terrain classification directly from Level-0 raw data and quantify the performance gap between in-wave computation via SIM-LIA and a fully digital DNN with comparable learnable parameters. The DNN uses two linear layers separated by a complex-valued ReLU activation. Unless otherwise stated, the SIM comprises $M=2048$ elements across $L=4$ transmissive layers with half-wavelength spacing ($d_x=d_y=\lambda/2$) and a total thickness of $0.05$\,m. The link employs a single transmit antenna and $K=4$ receive antennas at $f_c=5$\,GHz over a 150\,km space-to-ground path.
We leverage Sentinel-1 Level-0 raw synthetic aperture radar (SAR) I/Q data downloaded from Copernicus, which preserves the original phase and amplitude information, but lacks standard radiometric and geometric corrections.  
To enhance the learnability of Level-0 SAR data, each sample is augmented with a global 90$^\circ$ phase rotation and embedded at the first layer of the SIM \cite{liu2025onboard}. 
Although the phase rotation itself is a linear operation, it effectively redistributes signal components across the complex domain, facilitating feature separation in raw data representations. 
An ablation study further indicates that a 90$^\circ$ rotation provides the most robust performance in the SIM-LIA setup considered.
This fixed phase rotation preserves the signal amplitude while redefining the global I/Q phase, enabling more effective subsequent in-wave propagation.  Meanwhile, more sophisticated augmentations, such as phase noise and multipath effects, can be directly compatible with the proposed framework and left for future investigations.

\subsection {Confusion Matrix on SIM-LIA and DNN}

Fig.~\ref{fig:case2_confmat} compares SIM-LIA with a digital DNN on the four-class task under matched learnable-parameter budgets. 
SIM-LIA achieves class-wise accuracies of 63.7\%, 61.9\%, 60.5\%, and 66.7\% for ocean, forest, desert, and urban classification, respectively, showing that linear in-propagation transforms can already extract useful terrain structure directly from raw Level-0 SAR. 
Some off-diagonal confusion persists due to both speckle and the expressivity limits of a purely linear stack. 
The DNN attains over 79\% accuracy across all classes, with most cross-class errors below 11\%, due to its nonlinear decision boundaries, at the cost of \textmu s--ms latency, high energy consumption, and numerous floating point operations (FLOPs). 
By contrast, SIM-LIA performs inference at nanosecond-scale propagation latency, with very low energy consumption and near-zero FLOPs, since computation is embedded in passive wave propagation.
Overall, the result illustrates an accuracy--cost trade-off: the DNN serves as the best performer, while SIM-LIA attains ultra-fast, energy-efficient analog inference.
 
 \subsection{Convergence Analysis under Different Numbers of Classes}
 Using the same setup, Fig.~\ref{fig:convergence} plots epoch-wise accuracy for twin-class ($K=2$), triple-class ($K=3$), and quad-class ($K=4$) tasks. Solid curves adopt the standard 90° rotation, while dashed curves ablate this operation. Phase rotation consistently yields higher $\Delta_{\rm Acc}$ across all tasks, with relative gains of $55.68\%$ for $K=2$, $43.65\%$ for $K=3$, and $46.37\%$ for $K=4$, highlighting its role in enhancing data diversity and mitigating speckle noise and phase ambiguity in raw SAR signals.
The initial performance drop, most evident in the twin-class task, arises from random phase initialization and purely linear operations, which cause unstable analog modulation in early epochs. As training proceeds, effective phase patterns are rapidly learned, leading to stable convergence.
These results also indicate that, while SIM-LIA performs well in communication-enhancement tasks \cite{an2025stacked}, its purely linear nature limits expressiveness in more complex inference scenarios, motivating future exploration of wave-guided nonlinear mechanisms or hybrid analog–digital architectures \cite{Gao2023SPNN,Sanjari2025NonlinearMetasurface}.

\section{Conclusions and Future Work}
 We have surveyed SIMs as a wave-domain computing platform for communication, sensing, and onboard inference, and introduced a taxonomy spanning system scenarios, objectives, and analog computational functions. We reviewed some studies to show how SIMs enable beamforming, transforms, and correlation during wave propagation, while case studies on diffractive inference illustrated reductions in digital complexity and latency through physical-layer computation.
 Despite encouraging progress, SIM research remains at an early stage, with key challenges in accurate physical modeling, robustness under imperfect CSI, and the lack of open datasets and secure transmission mechanisms. Future work should focus on computation-aware physical models, scalable low-overhead control, and hybrid or nonlinear architectures, alongside collaborative efforts on datasets, benchmarks, and prototypes to establish SIMs as practical enablers of integrated wave-based communication, sensing, and computation schemes.
 
\bibliographystyle{IEEEtran}
 
\bibliography{myref}

\end{document}